\documentclass{eas}
\usepackage{ifpdf}
\usepackage{graphicx}
\begin{document}

\title{R$\&$D Status of Nuclear Emulsion for Directional Dark Matter Search} 
\author{T.Naka} \address{Institute for Advanced Research, Nagoya University,Japan}
\author{M.Kimura}\address{Department of physics, Toho University, Japan}
\author{M.Nakamura$^3$, O.Sato$^3$, T.Nakano$^3$,T.Asada}\address{Department of physics, Nagoya University, Japan}
\author{Y.Tawara}\address{ Division of Energy Science, EcoTopia Science Institute, Nagoya University, Japan}
\author{Y.Suzuki}\address{SPring-8, Japan Synchrotron Radiation Research Institute (JASRI), Japan }

\begin{abstract}

 In this study, we are doing R$\&$D for directional dark matter search with nuclear emulsion. First of all, higher resolution nuclear emulsion with fine silver halide crystals was developed in the production facility of emulsion at Nagoya university, and we confirmed that it can detect the expected nuclear recoil tracks. The readout of submicron tracks was required the new technology. We developed the expansion technique, and could readout the signal by shape analysis with optical microscopy. The two dimensional angular resolution is 36 degrees at the original track length of range from 150nm to 200nm with optical microscopy. Finally we demonstrated by using recoiled nuclei induced by 14.8MeV neutron, and confirmed the technique.Moreover, we developed the X-ray microscope system with SPring-8 as final check with higher resolution of selected candidate tracks with optical microscopy. The angular resolution was improved from 31degree with optical microscopy to 17degree with X-ray microscopy at the track length of range from 150nm to 250nm.  
We are developing the practical system and planning for start of the test running with prototype detector.

\end{abstract}
\maketitle
\section{Introduction}
 For directional dark matter search, detection of nuclear recoil tracks is essential technique (Spergel, 1988). As nuclear recoil energy is expected to be keV order, track length of recoiled nuclei becomes very short. Especially usual detectors with solid or liquid are very difficult to distinguish the tracks for the energy region.  However we proposed to use the nuclear emulsion which is a solid tracking detector as the directional dark matter detector. As nuclear emulsion has the highest spatial resolution, tracking of recoiled nuclei is possible in principle. We aim for directional dark matter search with large target mass.  

Nuclear emulsion is a kind of photographic film, and it can detect the 3D charged particle tracks. In present, nuclear emulsion is used in neutrino experiment, nuclear physics and other application. Target mass of nuclear emulsion in status big experiment is about 30000kg (N.Agafonova $\etal$, 2010).

By using nuclear emulsion, we proposed the probability of directional dark matter search. 
However this experiment with nuclear emulsion is required the different techniques from usual nuclear emulsion techniques. The essential issue is a shortness of tracks to detect. 
In this study, we proposed the new concept, and developed the new technique for nuclear emulsion itself and readout system.

\section{Fine grain nuclear emulsion for dark matter search}
Silver halide crystals dispersed in gelatin constitute the nuclear emulsion, and they are essentially important because depending on resolution and sensitivity.
The resolution of nuclear emulsion depends on silver halide crystal size and density. For a usual nuclear emulsion, crystal size is about 200nm, and line density of silver halide crystal that incident particles can penetrate is 2.3AgBr/${\mu}$m. This resolution is not enough for directional dark matter detection because expected track length of recoiled nuclei is less than 1$\mu$m in nuclear emulsion.

\begin{figure}[!h]
 \centering
\includegraphics[width=10cm,clip]
{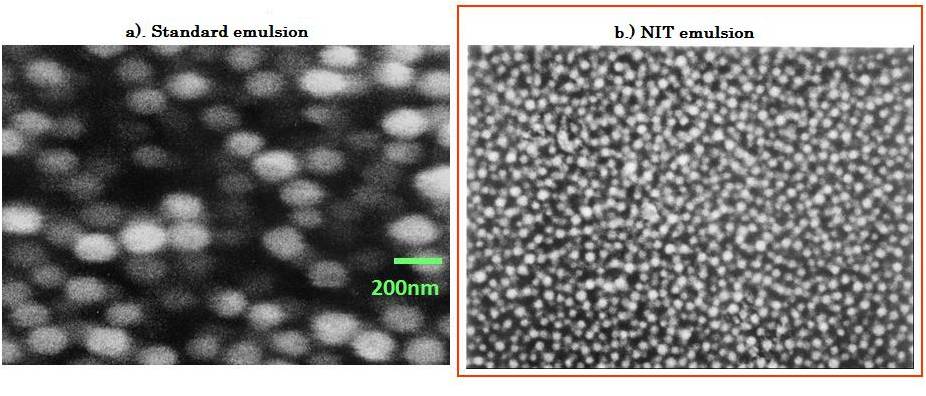}
\caption{Electron microscope images of nuclear emulsion. Figure a) shows the usual nuclear emulsion, figure b) shows the NIT produced by Fuji Film. }
\label{fig:Fig1}
\end{figure}

First of all, to resolve the this issue, we developed the fine grain nuclear emulsion with silver halide crystal of 40nm by collaboration with Fuji Film. This emulsion is called "Nano Imaging Tracker:NIT" (M.Natsume $\etal$,2007) and the resolution is five times as high as usual emulsion.  
Now, emulsion production became possible in Nagoya university by construction of production facility and collaboration with OB engineer of Fuji Film. Therefore, we are now possible to do a freely production and R$\&$D of emulsion. In Fig.\ref{fig:Fig1}, electron microscope images of usual emulsion and NIT emulsion are shown. 

\begin{figure}[!h]
 \centering
\includegraphics[width=10cm,clip]
{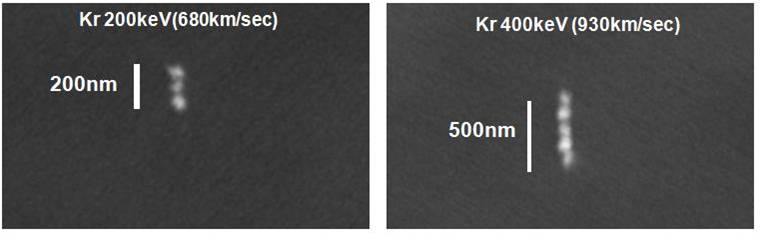} 
\caption{Scanning electron microscope images of Kr ion tracks. Left image shows the 200keV Kr ion and right image shows the 400keV Kr ion. }
\label{fig:Fig2}
\end{figure}

In production facility of our group, for the size of silver halide crystal which decides the spatial resolution, we have already produced the crystal size of 35$\pm$7nm stably. The ability for detection of nuclear recoil track can be recognized by low velocity ion with ion implant system. Here, we used low velocity Kr ion. For the emulsion produced by our group, submicron tracks of low velocity Kr ions could be distinguished by scanning electron microscope (SEM) like Fig.\ref{fig:Fig2}. Threshold of track length is about 100nm.  
 
Our immediate subject for R$\&$D of nuclear emulsion is sensitivity control and development of finer grain emulsion. The sensitivity of nuclear emulsion can be controlled by various method. Optimization of sensitivity is very important for high efficiency and background rejection. Development of finer grain emulsion will be also very important for lower threshold of detection.

\section{Concept for readout method of nuclear recoil tracks}
For detection of nuclear recoil tracks with nuclear emulsion, new readout method was required because expected track length is less than submicron.
 
Readout for search of nuclear recoil tracks should be used the optical microscope because large volume scanning is needed, and we have the technique of high speed readout stage of optical microscope. Although optical resolution is not enough to distinguish and identify the signal, we developed the new technique that can select the candidate tracks with optical microscope in this study. 
 
First, for nuclear emulsion after development treatment, as gelatin constitutes most of it, it can be expanded by swell characteristic. As silver grains constituting the track are binded by the polymer of gelatin, tracks are elongated with expansion of  emulsion. Here, ideal resolution of optical microscope is about 200nm at wavelength of 500nm and NA of 1.25. Therefore by two times expansion of emulsion, signal selection with optical microscopy is possible. By using chemical treatment, two times expansion is possible, and you can close to the ideal resolution by epi-illuminated dark field optics .
 
First of all, distortion of nuclear emulsion film for this technique should be checked. This 
is possible to measure the angular distribution of Kr ion tracks with SEM directly. For this observation, distortion of angle was less than 5.2degrees, where we assumed  $\sigma$ (standard deviation)  of scattering of 17degrees. by MC simulation.   
 Simultaneously, by measurement of track lengths with SEM observation, expansion rate was evaluated. For 400keV Kr ion tracks, mean track length before expansion was 182nm, and  the case of after expansion was 377nm. Therefore expansion rate is a factor of two.

\begin{figure}[!h]
 \centering
\includegraphics[width=10cm,clip]
{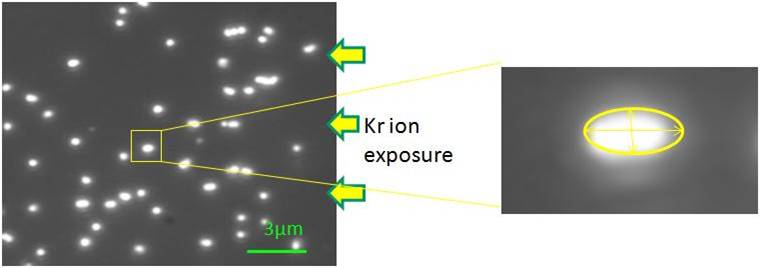}
\caption{Expanded Kr ion tracks with optical microscope. Left figure shows the area exists the elongated tracks. Right figure shows the concept of algorithm for readout of candidate tracks.}
\label{fig:Expanded_Kr_Tracks_Opt}
\end{figure}

We demonstrated the above concept by low velocity Kr ions. Optical microscope image of expanded submicron tracks were showed in Fig.\ref{fig:Expanded_Kr_Tracks_Opt}. Track in nuclear emulsion has a line of silver grains of more than two. Although each silver grains can not be distinguished on the optical resolution, nuclear recoil tracks can be selected by configuration of signal.

\begin{figure}[!h]
\centering
\includegraphics[width=10cm,clip]
{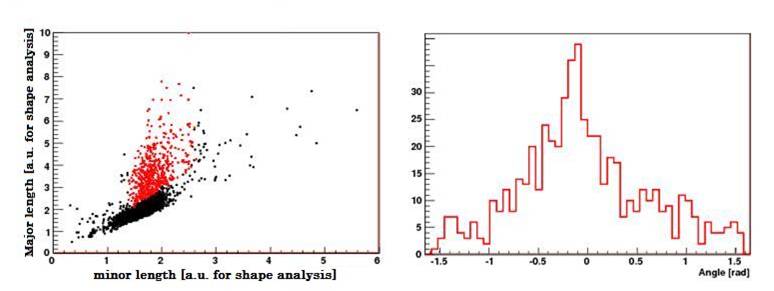}
\caption{The results of demonstration with Kr ion tracks for readout with optical microscopy. Left figure shows the distribution between major and minor lengths with elliptical fitting of shape analysis, where red points are candidate signals with temporary signal region cut. Right figure shows the angular distribution for candidates in signal region of left figure. }
\label{fig:Expanded_Kr_Tracks_Data}
\end{figure}%

For example, shape recognition as ellipse like Fig.\ref{fig:Expanded_Kr_Tracks_Opt} right can discriminate the candidate tracks well. By elliptical fitting of image processing, you can draw the distribution between major and minor length like Fig.\ref{fig:Expanded_Kr_Tracks_Data} left. Here we set the red area as temporary signal region (but reasonable). Another noises (random fogs) which are not tracks are ellipticity of one. Track candidates are expected to have ellipticity of more than one. Moreover angular distribution of Kr ion tracks can be measured automatically by this method. Angular distribution of Kr ion tracks was shown in Fig.\ref{fig:Expanded_Kr_Tracks_Data} right. Another , where 0rad. is direction of Kr ion tracks. 
From this, angular resolution of readout with optical microscopy is estimated as less than 36degrees, and immediate distinguishable track length with optical microscopy was 300nm. This track length corresponds to original track length of 150nm because expansion rate is a factor of two.
 
Finally, selected events can be confirmed by higher resolution microscopy. This tool is a X-ray microscope. It will be reported in Section.5.

\section{Demonstration using Neutron}
 As a demonstration of tracking for nuclear recoil events, we used the neutron of 14.8MeV induced by D-T nuclear fission. In this demonstration, target nuclei  assumed  Br or C,N,O from relationship between recoil energy and track length, but Ag nuclei are assumed to be not recognized because expected range is less than threshold of detector.

\begin{figure}[!h]
 \centering
\includegraphics[width=8cm,clip]
{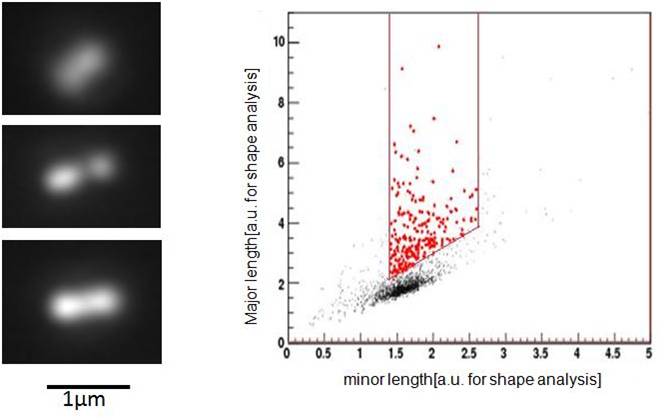}
\caption{The result of neutron recoil events. Left figures are the candidate recoiled tracks with optical microscopy. Right figure shows the distribution between major and minor length for elliptical shape analysis, where the red dots in red line are candidate tracks for temporary signal region cut. }
\label{fig:Neutron_Optical}
\end{figure}

Optical microscope images of detected candidate tracks were shown in Fig.\ref{fig:Neutron_Optical}. For this demonstration, candidate event density was 0.69$\pm$0.18/view. Here, one view is optical microscope view (10$\times$10$\times$1$\mu m$$^3$). As prediction by Geant4 simulation is 0.69/view at track length threshold of 100nm, status threshold of tracking is expected to be 100nm. However, predicted event density is consistent with the range from 100nm to 200nm in the allowable margin of error. This should be checked by higher resolution microscopy directly. I will report about this in Section.5.  

\begin{figure}[!h]
 \centering
\includegraphics[width=6cm,clip]
{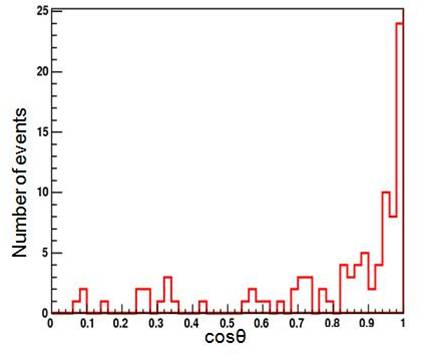} 
\caption{Angular distribution of neutron recoiled tracks with readout of optical microscopy. }
\label{fig:Neutron_AngularDist}
\end{figure}%

The result of detected candidate tracks induced by neutron was shown in Fig.\ref{fig:Neutron_Optical}. In right figure of Fig.\ref{fig:Neutron_Optical}, distribution between major and minor lengths of elliptical fitting for detected events was shown. Here, red area is the signal region for this study. The angular distribution of cosine for the events in signal region was shown in Fig.\ref{fig:Neutron_AngularDist}. We could detect the directionality for nuclear recoil events induced by neutron because cos$\theta$=1 is the direction of incidence of neutron. Angular dispersion (standard deviation) is 47$\pm$10degrees. On other hands, it was 54$\pm$2degrees on the Geant4 simulation. Hence angular dispersion is consistent with simulation.


\section{Confirmation of X-ray microscope system}
 Selected candidate tracks should be checked in higher resolution microscopy. Standard tool is electron microscope, but if you use it, destructive observation is inevitable. Hence, electron microscope is not available. 

 Optimal tool is the X-ray microscope (Akihisa Takeuchi $\etal$, 2009). Advantages of X-ray microscopy are as follow. 
 \begin{itemize} 
\item Higher resolution than optical microscopy. Effective resolution is 50nm.
\item Observation of thick sample is possible. At least, thickness of 100$\mu$m is possible to observe. 
\item No restriction of configuration. No-destructive observation is possible. 
\end{itemize}

As a prototype of X-ray microscope system, X-ray line of BL47XU at SPring-8,Japan was used. Here, X-ray energy was 8keV, and high contrast image was realized by using a phase difference method. 

Concept of this system is to check the candidate tracks selected by optical microscope in detail, then common coordinate between X-ray and optical microscope system is necessary. This can be achieved by photo mask pattern because you can decide the original point and Affine parameter. 

\begin{figure}[!h]
 \centering
\includegraphics[width=7cm,clip]
{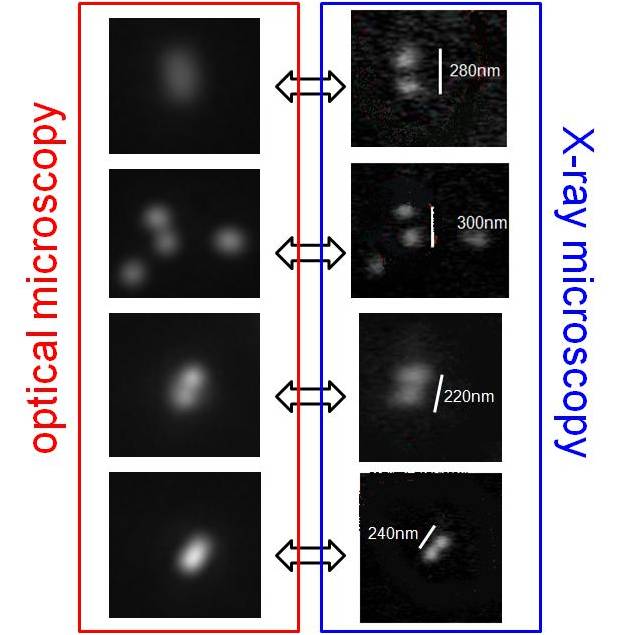} 
\caption{The result of neutron recoiled tracks which are made a comparison between optical and X-ray microscopy.}
\label{fig:Opt_and_Xray_matching}
\end{figure}

The result for X-ray microscope system was shown in Fig.\ref{fig:Opt_and_Xray_matching}. Here, left images are neutron recoil tracks with optical microscopy, and right images are matched tracks to optical microscopy with X-ray microscope.   

 We could make a comparison of the nuclear recoil tracks between optical and X-ray microscopy, and was possible the analysis the candidate tracks with high resolution. Here, for angular resolution was 31.4$\pm$4.7degrees with optical microscopy, it was 16.8$\pm$2.9 degrees with X-ray microscopy. By observation of X-ray microscope, angular accuracy can be improved because image of track blurs with scattering of light and diffraction limit for optical microscopy . 

By analysis with track by track, we confirmed that status readout system with optical microscope can distinguish the signal which track length is more than 150nm.

\section{Prospect}
 We are developing the practical tool and optimize the nuclear emulsion. 

By the development of optical microscope stage for readout of candidate tracks, large statistics analysis becomes available, and by optimization and tuning of optical condition, we have to decide the condition of signal selection and optimize the algorithm for selection of candidate events. In future, we will construct the high speed scanning system with GPGPU and high vision optical system. 

Simultaneously, we are developing the nuclear emulsion itself. Especially, by the nuclear emulsion with synthetic polymer, we control the background and crystal size for low threshold. In addition, we also optimize the development condition for high S/N and the control of configuration of silver grains. 
  
\section{Summery}
 We progress the R$\&$D of nuclear emulsion for directional dark matter search. For the experiment, nuclear emulsion with high resolution is required because tack length of recoiled nuclei is expected to be less than 1$\mu$m. We constructed the emulsion production facility, and could produce the fine grain nuclear emulsion which can detect the tracks of more than 100nm. This emulsion has higher quality than one produced by company. 

By using expansion technique, we confirmed that nuclear recoil tracks could be readout automatically. Moreover it was demonstrated by recoiled tracks induced by neutron. Finally we indicated that X-ray microscope function as a final check tool of candidate event selected by optical microscope with above technique. 

We are constructing the practical system, and planning the start of test running.


\end{document}